\documentclass[a4paper,11pt]{amsart}
\begin{document}
\hyphenation{gra-vi-ta-tio-nal re-la-ti-vi-ty Gaus-sian
re-fe-ren-ce re-la-ti-ve gra-vi-ta-tion Schwarz-schild
ac-cor-dingly gra-vi-ta-tio-nal-ly re-la-ti-vi-stic pro-du-cing
de-ri-va-ti-ve ge-ne-ral ex-pli-citly des-cri-bed ma-the-ma-ti-cal
de-si-gnan-do-si coe-ren-za pro-blem}
\title[Gravitational collapses to bodies of a finite volume]
{{\bf  Gravitational collapses\\to bodies of a finite volume}}

\author[Tiziana Marsico]{Tiziana Marsico}
\address{Liceo Classico ``G. Berchet'', Via della Commenda, 26 - 20122 Milano (Italy)}
\author[Angelo Loinger]{Angelo Loinger}
\address{Dipartimento di Fisica, Universit\`a di Milano, Via
Celoria, 16 - 20133 Milano (Italy)} \email{martiz64@libero.it}
\email{angelo.loinger@mi.infn.it}
\thanks{To be published on \emph{Spacetime \& Substance.}}

\begin{abstract}
We prove with an \emph{exact} relativistic computation that the
spherosymmetric gravitational collapses with a time-dependent
pressure end in bodies with a small, but \emph{finite} volume.
Against a diffuse, wrong conviction.
\end{abstract}

\maketitle


\vskip0.50cm \noindent \small \emph{\textbf{Summary}}. --
\textbf{1}. The recent discovery of the bright quasar HE
0450-2958; \emph{et cetera}. -- \textbf{2}. Relativistic
significance of a time-dependent hydrodynamical pressure.
\linebreak[4] -- \textbf{3}. Comments about papers by McVittie,
May and White, Chandrasekhar.\linebreak[4] -- \textbf{4}.
\emph{Exact} computation of a gravitational collapse with a
time-dependent pressure. -- \textbf{4bis}. A remark on the
polytropic equations of state. -- \textbf{5}. External forms of
solution and the problem of continuity adjustments between
internal and external solutions. -- \textbf{6}. Origin of a double
error by Chandrasekhar and followers. -- Appendix: The singular
\emph{loci} as defective points of any continuum theory; \emph{et
cetera}.

\normalsize

\vskip1.20cm \noindent \textbf{1}. -- The fictitiousness of the
notion of black hole (BH) has been demonstrated with many
arguments \cite{1}. However, the majority of the astrophysicists
are still convinced of the real existence of the BH's, in
particular of the supermassive BH's: thus, ``A quasar is thought
to be powered by the infall of matter onto a massive black hole at
the centre of a massive galaxy'' \cite{2}. (See also \cite{2bis}).
In reality, \emph{all} the claimed observational discoveries of
BH's show only the existence of very large, or enormous, masses
concentrated in relatively small volumes (see, e.g., \cite{3}).
The discovery of the bright quasar HE 0450-2958 (some 5 billion
light-years away), which does \emph{not} reside in the centre of a
massive host galaxy, has generated some interpretative difficulty.
According to an explanation of the authors of the paper quoted in
\cite{2}, ``the black hole of HE 0450-2958 lies in a galaxy with
not only a stellar content much lower than average, but also with
a dark halo $\ldots$''.

\par In the present paper we give a \emph{new} argument against
the physical reality of the BH's: we show that a sufficiently
realistic computation of the relativistic gravitational collapse
of a massive (or supermassive) object yields as a final product a
body with a small, but \textbf{\emph{finite}} volume. We infer, in
particular, that the quasar phenomenon is a consequence of a
permanent inward mass flow of a galactic origin, which accumulates
onto a supermassive body restricted within a \emph{finite}, small
volume.

\par An important by-product of our computation is the following:
our internal solution confirms a general fact (see \cite{4} and
references therein), which contradicts a widespread, naive
opinion: the accelerated motions of the matter particles of  our
collapsing body do \emph{not} generate gravitational waves (GW's);
these particles describe geodesic lines. \cite{5}.

\par A last remark, which concerns the concept of gravitational
potential energy in general relativity (GR): this concept is
extraneous to the spirit of GR, for the simple reason that it
cannot be formulated in a tensorial manner. Here a (partial)
analogy with Poynting theorem of Maxwell electrodynamics is
enlightening: the energy of a closed system of charged particles
that are in motion under the action of their e.m. fields is
composed of two parts: the total kinetic energy of the charges
plus the energy of the global e.m. field; the consideration of the
instantaneous Coulombian interactions between the particles
destroys the \emph{manifest} Lorentzian covariance. Nevertheless,
the Newtonian concept of gravitational potential energy is
practically useful -- and theoretically sensible -- in several
astrophysical instances: e.g., if we assume, with a good
approximation,  that the potential energy of any spherical mass
$M$ of radius $R$ is $GM^{2}/R$ (where $G$ is the gravitational
constant), we have an energy of $10^{60}$ erg when $M=10^{8}$
solar masses and $R=3.8\times10^{4}$ solar radii, with an average
density of $2.57 \times 10^{-6}$ g/cm$^{3}$; the same potential
energy is obtained for $M=10^{6}$ solar masses and $R=3.8$ solar
radii, with a density of $2.57\times10^{4}$ g/cm$^{3}$ (see
further \cite{6}).

\vskip0.50cm \noindent \textbf{2}. -- With standard notations, the
(pre-relativistic) Euler equation of perfect fluids and the
continuity equation are, as it is well known
$[\mu=\mu(\textbf{r},t),\, p=p(\textbf{r},t)]$:

\begin{equation} \label{eq:one}
\mu \frac{\textrm{d}\textbf{v}}{\textrm{d}t} = \mu \,
\textrm{\textbf{F}} - \textrm{grad}\,p \quad; \quad (\mu=f(p),
\textrm{equation of state}) \quad,
\end{equation}

\begin{equation} \label{eq:two}
\frac{\textrm{d} \mu}{\textrm{d}t} + \textrm{div}\,(\mu \,
\textbf{v})=0 \quad;
\end{equation}

a mere time dependence of pressure, $p=p(t)$, is here of a scarce
interest. On the contrary, in the theory of relativity such time
dependence can have important physical consequences -- as we shall
see in the sequel. For the time being, we limit ourselves to
remember that in special relativity (SR) we have the following
Eulerian equations $(j=0,1,2,3)$; $u^{j}$ is the four-velocity:

\begin{equation} \label{eq:three}
\left(\mu+\frac{p}{c^{2}}\right)\,c\,\frac{\textrm{d}u^{j}}{\textrm{d}s}
= \mu \,F^{j}+ \frac{\partial p}{\partial x_{j}} - \frac{1}{c}\,
\frac{\textrm{d}p}{\textrm{d}s}\, u^{j}\quad ; \quad
[\mu=\varphi(p)] \quad ;
\end{equation}

eqs.(\ref{eq:three}) follow from $\partial T^{jk}/\partial
x^{k}=\mu F^{j}$, with

\begin{equation} \label{eq:threeprime} \tag{3'}
c^{2}\, T^{jk}= \left(\mu+\frac{p}{c^{2}}\right) u^{j}u^{k}-p\,
\eta^{jk}\quad,
\end{equation}

where $\eta^{jk}$ is the customary Minkowskian metrical tensor.

\vskip0.50cm \noindent \textbf{3}. -- An interesting essay on the
gravitational collapses of massive celestial bodies of a spherical
shape has been written by McVittie in 1964 \cite{6}. Subsequent
authors have strangely ignored his results. It is here sufficient
to cite a paper by May and White (1966) \cite{7} and a review
article by Chandrasekhar (1972) \cite{8}, because their erroneous
conclusion that any relativistic gravitational collapse of a
spherical body ends always in a BH has become a \emph{caput
doctrinae sacrae} for the astrophysical community.

\par In a previous paper \cite{1} we have proved that an
appropriate treatment of relativistic gravitational collapses with
\emph{zero} pressure yields finally \emph{a point mass}, \emph{not
a BH}. By revisiting and completing a result by McVittie \cite{6},
we shall show that relativistic gravitational collapses of
spherical bodies with pressure and density which are
\emph{functions of the time alone} end in objects of a
\textbf{\emph{finite}} volume. It is evident that a computation
with pressure and density which are functions of time \emph{and}
of radial co-ordinate would strengthen this result. It contradicts
the statement of Chandrasekhar \cite{8} that in GR the allowance
for pressure does not prevent the matter from collapsing to a BH.
We shall see in sect.\textbf{6}  the reason of this
Chandrasekhar's error, which is really a double error.

\vskip0.50cm \noindent \textbf{4}. -- We adopt notations (with
some inessential modification), units of measure (CGS-system), and
internal reference frame as in \cite{6}. The problem: to study a
relativistic gravitational collapse with spherical symmetry under
the assumption that matter density and pressure are \emph{only}
time-dependent.

\par In a Gaussian-normal (or synchronous) and co-moving reference
system, we can write, assuming -- for simplicity's sake only --
that the constant spatial curvature $\kappa$ is equal to
\emph{zero} (FRW metric with $\kappa=0$):

\begin{equation} \label{eq:four}
\textrm{d}s^{2}= c^{2}\textrm{d}t^{2} - S^{2}(\tau)\
[\textrm{d}r^{2}+r^{2}\textrm{d}\Omega^{2}] \quad,
\end{equation}

\begin{equation} \label{eq:five}
\textrm{d}\Omega^{2}:= \textrm{d}\vartheta^{2}+\sin^{2}\vartheta
\, \textrm{d}\varphi^{2} \quad,
\end{equation}

where: $\tau:=t/T$, and $T$ is a fixed time interval, whose
physical meaning will be clear presently; the function $S(\tau)$
will be determined as a solution of Einstein field equations. We
assume that at the initial instant ($t=0$) of the collapse $S(0)
\equiv S_{0}=1$.

\par The celestial material is a perfect fluid of density
$\varrho(\tau)$ and pressure $p(\tau)$. The only components of
mass tensor $T_{jk}$, $(j,k=0,1,2,3)$, which are different from
zero, are the following:

\begin{equation} \label{eq:six}
T^{00}= \varrho\,(\tau) \quad, \quad
S^{2}(\tau)\,T^{11}=S^{2}(\tau)\,r^{2}\,T^{22}=S^{2}(\tau)\,r^{2}\sin^{2}\vartheta
\,T^{33}=p(\tau) \quad.
\end{equation}

Let us denote with a subscript a derivative with respect to
$\tau$. The Einstein equations give:

\begin{equation} \label{eq:seven}
8\pi \,G\,\varrho \, T^{2}= 3\,\frac{S_{\tau}^{2}}{S^{2}} \quad,
\end{equation}

\begin{equation} \label{eq:eight}
8\pi \,G\, \frac{p}{c^{2}}\,T^{2}= -2\,\frac{S_{\tau\tau}}{S}-
\frac{S_{\tau}^{2}}{S^{2}} \quad;
\end{equation}

the four equations $\{$covariant divergence of mass tensor equal
to zero$\}$ yield only:

\begin{equation} \label{eq:nine}
\frac{\textrm{d}}{\textrm{d}\tau} (\varrho S^{2}) +
\frac{p}{c^{2}} \frac{\textrm{d}S^{2}}{\textrm{d}\tau}=0 \quad.
\end{equation}

It follows from eq.(\ref{eq:four}) that the volume of our
spherical body is equal to $(4/3)\,\pi\,[S(\tau)\, r_{b}]^{3}$,
where $r_{b}$ is the radial co-ordinate of the boundary; since
this volume is contracting, the derivative $S_{\tau}(\tau)$ is
negative. Accordingly, from eq.(\ref{eq:seven}) we obtain

\begin{equation} \label{eq:ten}
(S_{\tau})_{0}=- \left(\frac{8}{3}\,\pi
\,G\,\varrho_{0}\right)^{1/2}T \quad,
\end{equation}

if $\varrho_{0}:=\varrho(0)$. Let us assume with McVittie the
following \emph{equation of state}:

\begin{equation} \label{eq:eleven}
\frac{p}{c^{2}} = \varepsilon_{0}\,\varrho_{0}
\left(\frac{\varrho}{\varrho_{0}}\right)^{3/2} \quad,
\end{equation}

where: $\varepsilon_{0}:=p_{0}/(c^{2}\varrho_{0})$; $p_{0}:=p(0)$.

\par Then, the solution of eqs.(\ref{eq:seven}) and
(\ref{eq:eight}) can be written as follows (see \cite{9}):

\begin{equation} \label{eq:twelve}
\varrho =
\varrho_{0}\,\left[\left(1+\varepsilon_{0}\right)S^{3/2}-\varepsilon_{0}\right]^{-2}
\quad,
\end{equation}

\begin{equation} \label{eq:thirteen}
\tau = \frac{1+\varepsilon_{0}}{(6\,\pi \,G\,\varrho_{0})^{1/2}T}
\left(1-S^{3/2}+\frac{3}{2}\,\frac{\varepsilon_{0}}{1+\varepsilon_{0}}\ln
S \right)\quad;
\end{equation}

the functions $S(\tau)$ and $\varrho(\tau)$ are thus determined.

\par At time $t=T$, i.e. when $\tau=1$, we have that
$\varrho(1):=\varrho_{fin}$ and $p(1):=p_{_{fin}}$ become
infinite, and

\begin{equation} \label{eq:fourteen}
S(1):=S_{fin}=\varepsilon_{0}^{^{2/3}}(1+\varepsilon_{0})^{-2/3}
\quad;
\end{equation}

consequently, at the final stage of the collapse \emph{our body
has a} \textbf{\emph{finite}} \emph{volume equal to} $(4/3)\,\pi
(S_{fin}r_{b})^{3}$.

\par By substituting $S_{fin}$ of eq.(\ref{eq:fourteen}) into
eq.(\ref{eq:thirteen}), we obtain the duration $T$ of the
collapse:

\begin{equation} \label{eq:fifteen}
T= \frac{2}{3} \left(1+\varepsilon_{0}
\ln\frac{\varepsilon_{0}}{1+\varepsilon_{0}} \right)
\left(\frac{8}{3}\,\pi \,G\,\varrho_{0} \right)^{-1/2} \quad.
\end{equation}

By differentiating eq.(\ref{eq:thirteen}):

\begin{equation} \label{eq:sixteen}
S_{\tau}^{2}= \frac{4}{9} \left(1+\varepsilon_{0}
\ln\frac{\varepsilon_{0}}{1+\varepsilon_{0}} \right)^{2}
S^{2}\,\left[\left(1+\varepsilon_{0}\right)S^{3/2}-\varepsilon_{0}\right]^{-2}
\quad;
\end{equation}

from which: $(S_{\tau})_{fin}=-\infty$.

\par For a \emph{zero}-pressure collpase, $\varepsilon_{0}=0$, and
eqs.(\ref{eq:fourteen}) and (\ref{eq:fifteen}) give:

\begin{equation} \label{eq:fourteenprime} \tag{14'}
[S_{fin}]_{p=0}=0 \quad,
\end{equation}

\begin{equation} \label{eq:fifteenprime} \tag{15'}
[T]_{p=0}=\frac{2}{3}(8\,\pi \,G\,\varrho_{0})^{-1/2} \quad,
\end{equation}

i.e. \emph{a} \textbf{\emph{zero}} \emph{final volume} and a
duration of collapse \emph{greater} than $T$.

\par From a conceptual standpoint, we could stop here: the above
(internal) solution is  completely exhaustive: if $p=p(t)\neq 0$
at the end of the collapse the body has a \emph{finite} volume.

\par However, it has become customary, in the investigations of
gravitational collapses, to study also the \emph{external}
solution of the problem, i.e. the gravitational potential $g_{jk}$
\emph{outside} of the collapsing object. But here McVittie's
approach met with an unexpected difficulty: the \emph{continuity
adjustment} of the two solutions (internal and external) at the
boundary of the material sphere succeeded  only for the
\emph{zero}-pressure collapses. In the more interesting case of a
\emph{non}-zero pressure $p=p(t)$ it failed. Our author wrote
\cite{6}: ``It may, of course, be true that other forms of
external solution exist $[\ldots$ that $]$ might insure the
continuity at the boundary $[$in all the considered instances$]$.
If such form exists, I am not aware that they have been
discovered.''

\vskip0.50cm \noindent \textbf{4bis}. -- Equation
(\ref{eq:eleven}) is the equation of state of a given polytropic
model of the celestial material. For our problem it is, of course,
a little schematic model. However, polytropic equations are not
infrequently adopted also for material densities above those of
nuclear matter, owing to the uncertainties concerning the real
equation of state in such cases.

\par We think that our final result -- i.e., the collapse to a
\emph{finite} volume -- is qualitatively independent of the
precise structure of the equation of state.

\vskip0.50cm \noindent \textbf{5}. -- A physically interesting and
\emph{general form} of external solution has been given by
Eddington at p.94 of his treatise \cite{10}:

\begin{equation} \label{eq:seventeen}
\textrm{d}s^{2}=\left[1-\frac{2m}{f(r)}
\right]c^{2}\textrm{d}t^{2} -  \left[1-\frac{2m}{f(r)}
\right]^{-1} [\textrm{d}f(r)]^{2} - [f(r)]^{2}\textrm{d}\Omega
\quad;
\end{equation}

here: $m\equiv GM/c^{2}$; $M$ is the mass of the spherical
distribution of matter; $f(r)$ is \emph{any} regular function of
the radial co-ordinate $r$.

\par Strictly speaking, eq.(\ref{eq:seventeen}) was derived by
Eddington under the assumption that the gravitational field is
generated by a point mass at rest. But it is obvious that, with
\emph{suitable} choices for $f(r)$, it holds also for the outside
of \emph{any} spherical distribution at rest-- or even  in any
spherosymmetrical motion, by virtue of a well-known Birkhoff's
theorem.

\par Examples. -- For a material point: if $f(r)\equiv r$, we have
the form of solution by Hilbert, Droste, Weyl, the so-called
\emph{standard} form (improperly denominated ``by
Schwarzschild''); for $f(r)\equiv [r^{3}+(2m)^{3}]^{1/3}$, we have
the \textbf{\emph{original}} Schwarzschild form \cite{11}; for
$f(r)\equiv r+2m$, we have a form which was first investigated by
M. Brillouin; \emph{etc}. \emph{etc}. For a sphere of an
incompressible fluid Schwarzschild \cite{12} found an internal and
an external solution; the latter can be formally obtained from
(\ref{eq:seventeen}) by putting $f(r)\equiv
[r^{3}+\delta^{3}]^{1/3}$, where the constant $\delta\neq 2m$ is
given by his formula (33). \emph{N}.\emph{B}.: the constant
$\delta$ -- denoted with $\varrho$ by Schwarzschild -- is
different from $2m$ as a consequence of the fact that, in perfect
analogy with Newton theory, He prescribed continuity conditions on
the boundary of the sphere \emph{both} for $g_{jk}$ \emph{and} for
$\partial g_{jk}/\partial x^{\alpha}$, $(\alpha=1,2,3)$. --

\par Owing to the ``flexibility'' of eq.(\ref{eq:seventeen}), it
is \emph{certain} that there exist functions $f(r)$ which are apt
to describe correctly the outside of any collapsing spherical body
for which eq.(\ref{eq:four}) is valid. However,
eq.(\ref{eq:seventeen})  is not written in a Gaussian-normal and
co-moving frame as eq.(\ref{eq:four}) is. Therefore, to verify the
existence (or non-existence) of continuity conditions on the
periphery of the sphere we should transform eq.(\ref{eq:four})
into the generic Eddington's reference system, or
\emph{vice-versa} eq.(\ref{eq:seventeen})  into a Gaussian-normal
form. To our aims, we can avoid this tedious procedure, the
following trivial argument being sufficient to rule out the
possibility of a collapse to a BH. Indeed, the values $S(\tau')$,
$S(\tau'')$, \dots of function $S(\tau)$ at different times $t',
t''$ $\ldots$ of the Gaussian frame of eq.(\ref{eq:four}) have
corresponding values, say $f(r[b'])$, $f(r[b''])$, $\ldots$ --
where $r[b']$, $r[b'']$, $\ldots$ are radial co-ordinates of the
spherical boundary \emph{b} -- in the reference system of
eq.(\ref{eq:seventeen}). In particular, $S_{fin}$ has a
corresponding value $f(r[b_{fin}])$ in Eddington's frame. Now, any
function $f(r)$ in eq.(\ref{eq:seventeen}), which does \emph{not}
originate an event horizon, excludes \emph{ipso facto} the fictive
notion of BH.

\par Even if in the present problem GR did not allow a continuity
adjustment, the above reasoning would remain adequate to our
purpose: to demonstrate the unreality of the BH's.

\vskip0.50cm \noindent \textbf{6}. -- We can see now the origin of
the double error by Chandrasekhar \cite{8}, i.e. of his assertion
that a non-zero pressure does not hinder the spherical body from
collapsing into a BH.

\par First of all, we observe that the original form of solution
given by Schwarzschild \cite{11} for a point mass (and which can
be formally obtained by putting $f(r)\equiv
[r^{3}+(2m)^{3}]^{^{1/3}}$ in eq.(\ref{eq:seventeen})) is
everywhere regular, with the only exception of the origin $r=0$:
there is here no room for the fantastic notion of BH. However, in
the relativistic literature has prevailed the HDW-standard form of
solution, for which $f(r)\equiv r$. The Schwarzschildian form is
diffeomorphic to standard form \emph{only if this form is
considered for} $r>2m$. This is \emph{no} physical restriction --
and only for $r>2m$ one can call (with admissible impropriety)
``by Schwarzschild'' the standard form. Obviously, the condition
$r>2m$ is sufficient to exclude the BH, whose fictitiousness can
be proved in several ways. The simplest one is as follows: the
existence of the BH's would demand that, in the internal region of
the singular surface $r=2m$, the radial co-ordinate $r$ (which was
chosen so that the area of the surface $r=\textrm{const}$ be $4\pi
(\textrm{const}^{2})$ became a time co-ordinate; and
\emph{vice-versa}: a mathematical and physical absurdity, as it
was stressed by Nathan Rosen. We remember finally that \emph{no}
physical result depends on the choice of the function $f(r)$ --
provided that, of course, we exclude the fictive singularities.

\par Chandrasekhar was a convinced believer in the real existence
of the BH's (first error), and this conviction had a momentous
consequence for the problem of a collapsing sphere, since he
wrote: ``There is no alternative to the matter collapsing to
infinite density at a singularity once a point of no return $[$an
event horizon, the surface $r=2m$ of the standard form of the
external solution$]$ is passed''. In this way, the final result of
the collapse would be the same as in a zero-pressure case (second
error). The two errors have a unique origin: the belief that the
standard form has a mathematical and physical meaning \emph{even
for} $r\leq 2m$. It seems that Chandrasekhar and the great
majority of the astrophysicists have not read, in particular, the
fundamental memoirs by Schwarzschild (\cite{11}, \cite{12}) and
Eddington's treatise \cite{10}. (A tentative psychological
explanation: perhaps Chandrasekhar and followers have been misled
by an improper interpretation of the usual version of Birkhoff's
theorem, i.e. by the assertion that the time-independent
\emph{standard} form is valid outside of \emph{any} spherical
distribution of matter. But they have forgotten that for the
Fathers of Relativity this form holds -- and for good reasons --
\emph{only for} $r>2m$ \cite{13}.

\par \textbf{\emph{In conclusion}}: according to Chandrasekhar and
followers, the final result of all spherical gravitational
collapses is always a BH, both for $p=0$ and for $p\neq 0$. On the
contrary, we have demonstrated (see \cite{1} and previous
sect.\textbf{4}) that when $p=0$ the final stage of collapse is
simply a \emph{material point}, not a BH; and when $p=p(t)\neq 0$
is a body of a small, but \emph{finite} volume (see
sect.\textbf{4}). \vskip0.30cm

\vskip0.30cm
\begin{center}
\noindent \small \emph{\textbf{APPENDIX}}
\end{center}

\normalsize \noindent \vskip0.20cm
\par With regard to the last term at the right side of
eqs.(\ref{eq:three}), Einstein emphasized that it is absent in
pre-relativistic hydrodynamics \cite{14}. However, its
contribution to gravitational collapses, although small, is
\emph{not} negligible, as we have seen.

\par It seems a little queer that Chandrasekhar and followers have
preferred to direct their attention on the various kinds of
singularities of Einsteinian fields \cite{15} rather than try to
solve precise problems on the gravitational collapses, as McVittie
did.

\par On the other hand, it is obvious that accurate results
proving that GR does not represent an overturning of Newton
theory, but a refinement and a completion of it, do not impress
the public opinion and the politicians, who decide the financing
of scientific researches. On the contrary, the belief in fictive
notions (as the BH's) and in physically unreal entities (as the
GW's) favours fanciful divagations that stir up popular curiosity
and \emph{\'epatent les bourgeouis}.

\par Immediately after the turning of tide, we shall assist at
amusing theatrical recriminations.

\small \vskip0.5cm\par\hfill {\emph{Wenn man f\"ur's K\"unftige
was erbaut,}
  \par\hfill \emph{Schief wird'ìs von vielen angeschaut.}
     \vskip0.10cm\par\hfill J.W. von Goethe}

\normalsize


\end{document}